# DEVOPS AND AGILE METHODS INTEGRATED SOFTWARE CONFIGURATION MANAGEMENT: HAVELSAN EXPERIENCE

Fatih Bildirici
HAVELSAN KKST R&D Group Directorate
06510 Ankara, Turkey

Asst. Prof. Keziban Seçkin Codal
Ankara Yıldırım Beyazıt University
06810 Ankara, Turkey

**Abstract**
The advancements in the software industry, along with the changing technologies, methods, and conditions, have particularly brought forth a perspective that prioritizes the improvement of all stages of the software development lifecycle by approaching the process through automation. In particular, methods such as agile methodologies and DevOps, which focus on collaboration, automation, and efficient software production, have become crucial for the software industry. In particular, the understanding of utilizing principles such as distribution management, collaboration, parallel development, and end-to-end automation in agile software development, and DevOps techniques has emerged. In this study, one of these areas, software configuration management, and the integration of modern software development practices such as agile and DevOps are addressed. The aim of this study is to examine the differences and benefits that innovative methods bring to the software configuration management field when compared to traditional methods. To this end, a project is taken as a basis, and with the integration of DevOps and agile methodologies, improvements are made and the results are compared with the previous state. As a result of monitoring software configuration management with the integration of DevOps and agile methodologies, improvements are seen in the build and deployment time, automated report generation, more accurate and fault-free version management, completely controlling the software system, working time and workforce efficiency.

**Özet**
Yazılım sektöründeki gelişmeler, değişen teknolojiler, yöntemler ve koşullar özellikle yazılım geliştirme yaşam döngüsü boyunca tüm aşamaların iyileştirilmesini sürecin otomatize edilerek ele alınmasını önceleyen bir perspektif ortaya çıkarmıştır. Özellikle çevik yöntem ve DevOps gibi işbirliğine, otomasyona, hızlı ve efektif yazılım üretimine odaklanan yöntemler yazılım sektörü için önem kazanmıştır. Özellikle dağıtım yönetimi, iş birliği, paralel geliştirme, uçtan uca otomasyon gibi prensiplerin, çevik yazılım geliştirme, DevOps gibi tekniklerin önem kazanması ve sürekli iyileştirme için kullanılması anlayışı ortaya çıkmıştır. Bu çalışmada bu alanlardan biri olan yazılım konfigürasyon yönetimi ile çevik ve DevOps gibi modern yazılım geliştirme uygulamalarının entegrasyonu ele alınmıştır. Çalışmanın amacı geleneksel metodlar ile karşılaştırıldığında yenilikçi yöntemlerin entegrasyonunu yazılım konfigürasyon yönetimi alanında ortaya çıkaracağı farkları ve faydaları ele almaktır. Bu kapsamda ele alınan bir proje yardımıyla DevOps ve çevik yöntem entegrasyonu ile otomatize iyileştirmeler yapılmadan önceki hali temel alınıp özetlenerek, sürekli iyileştirme motivasyonuyla ilgili düzenlemeler yapılmış ve sonuçlar karşılaştırılmıştır. DevOps ve çevik yöntem entegrasyonu ile yazılım konfigürasyon yönteminin takibi sonucunda, inşa ve yükleme süresinde, otomatize rapor üretmede, versiyon yönetiminin daha doğru ve hataya kapalı hale getirilmesinde, yazılım sisteminin tamamen kontrol altında tutulmasında, çalışma süresinde ve işgücü verimliliğinde iyileşmeler görülmüştür.
**Anahtar Kelimeler:** DevOps, Çevik Yöntem, Yazılım Konfigürasyon Yönetimi, Yazılım Geliştirme Yaşam Döngüsü

# 1. Problem Definition, Aim of the Study and Related Studies

Software configuration management is important for software companies such as HAVELSAN operating in sectors with regulated structures and standards such as defence industry. However, as we focus on in this study, traditional software configuration methods can be slow, rigid and burdensome. This structure and the problems that led to our study can be expressed as follows:

- Lack of agility and long delays in implementing updates or changes, responding to changing situations
- Limited co-operation and feedback structure, making it difficult to identify and address problems in a timely manner
- Lack of automation in the configuration management process, resulting in manual errors and the need for increased time and resources

The aim of the study is to define a structure that can basically respond to these problems and to emphasise the gains obtained in a sample project where this structure is applied. In addition, the primary aim of the study is to provide an infrastructure for models that can be developed and made more efficient and effective with new studies.

Recent research has shown the potential benefits of integrating agile and DevOps methodologies for software configuration management. For example, a study by (Durrani, Pita, Richardson, & Lenarcic, 2014) found that using an agile approach to configuration management leads to faster and more efficient implementation of updates and changes. Another study by (Hochbergs & Nilsson Sjödahl, 2020) found that using DevOps principles in configuration management improves the speed and flexibility of the process while increasing collaboration and communication between teams.

Other studies have also emphasised the importance of automation and collaboration in configuration management. A study by (Ebert, Gallardo, Hernantes, & Serrano, 2016) found that automation in the configuration management process increases efficiency and reduces the risk of errors. Another study by (Bendix & Ekman, 2009) emphasised the importance of collaboration and communication in ensuring the success of configuration management, especially in large and complex systems.

These studies show the potential benefits of integrating agile and DevOps methodologies for software configuration management also in a regulated environment. However, further research is needed to fully understand the impact of this approach and to identify best practices for its implementation.

## 2. Introduction

Software configuration management is an approach that increases the stability and reliability of systems by improving software development processes (Berczuk, Berczuk, & Appleton, 2003). Its purpose is to monitor and control changes made during a software development project. This includes managing source code, documentation and other related files, and properly documenting, reviewing and approving changes to these files. In addition, configuration management involves planning, controlling and coordinating the software operating environment and overall configuration (Berczuk, Berczuk, & Appleton, 2003).

In the context of software development, configuration management also refers to the process of configuring and maintaining all environments that host the software. By applying software configuration management, it is possible to ensure that the software is properly configured, versions and changes are managed, the developed software works as intended, and the development process is designed and implemented accordingly. This approach has recently become more efficient and effective with the integration of DevOps methods and processes and the Agile method (Bendix & Ekman, 2009). Organisations' use and integration of the right software configuration management processes and tools contribute to this efficiency (Moreira, 2010). Examples of these strategies include Gitflow strategies, automation and integration approaches, collaboration and communication methods. By integrating DevOps, Agile and software configuration management, it is possible to respond quickly to changing requirements and needs while delivering high quality software that provides value to customers (Hochbergs & Nilsson Sjödahl, 2020). This approach has been implemented at HAVELSAN as an important element of effective system and software development and an approach that is compatible with software excellence certifications and standards has been introduced. In this study, we provide a perspective on the theoretical framework of this new approach and present the framework of this experience and the implementation of software configuration management in HAVELSAN, integrated with DevOps and agile methods, strengthened with new technologies and approaches.

## 3. DevOps and Agile Integrated Software Configuration Management

The integration of DevOps, agile methods and Software Configuration Management (SCM) is crucial for the successful development and deployment of complex systems. DevOps is a collaborative and automation-based approach to software development that aims to optimise the entire software delivery process, including development, deployment and operations (Hemon, Lyonnet, Rowe, & Fitzgerald, 2020). Agile methods can be summarised as a flexible, iterative approach that focuses on delivering high quality, valuable software to customers quickly and efficiently (Laukkarinen, Kuusinen, & Mikkonen, 2018). Software Configuration Management, as mentioned, is a discipline that controls and manages all elements from infrastructure to code, from documentation to configuration in order to define software elements, manage changes and versions, and create controlled and stable software throughout the software development life cycle (Zimmermann & Bird, 2017).

Especially in complex systems, software configuration management is very important to ensure that the various components of the system are properly integrated and that changes made in one component do not adversely affect other components. Integrating software configuration management with DevOps and agile methods enables organisations to respond quickly to changing customer needs and market conditions while delivering high quality software that provides value to customers (Hochbergs & Nilsson Sjödahl, 2020). While this integration increases collaboration and communication for teams, the power of automation and the contribution of new technologies to efficiency enable faster and testable responses to customer needs and requirements and rapid reflection of changes under control.

Combining and integrating software configuration management, DevOps and agile methods offers another important practical advantage. The principles of teamwork, release management, version control and change management are all aligned with the principles of continuous monitoring, automation and continuous integration emphasised in DevOps and agile methods (Hochbergs & Nilsson Sjödahl, 2020). This holistic approach can be applied at different stages in software systems, resulting in a process suitable for achieving an overall improvement (Yarlagadda, 2021).

## 4. HAVELSAN Experience

HAVELSAN is a defence industry company specialised in defence, automation, simulation, information and communication technologies, cyber security and software development. The company, which operates mainly on a project basis with over 2000 qualified employees, focuses on critical success factors such as efficiency, innovation and continuous improvement in software development and R&D activities as well as all its projects. In this study, the application of DevOps and Agile principles integrated with software configuration management as a means of improving software systems and maturing the software development process within HAVELSAN is analysed.

HAVELSAN has a rigorous process for planning and gathering requirements for development projects, especially for complex and mission critical software and defence software. In this context, system, hardware and software configuration management is a very important part of the planning and implementation process. The model developed for configuration management in software and application development aims to improve the process by using modern development techniques, to enhance software configuration management and to better meet the basic configuration criteria. This approach includes identification, stabilisation for development and improvement, management of changes, version control and continuous improvement. It also includes ensuring the suitability of the development process for parallel development, faster and more stable release and managing changes, release, environment, configuration. In order to achieve this goal, the relevant model was applied on a project basis and the results were monitored.

Prior to the implementation of the software configuration management (SCM) approach, which integrates DevOps and agile methods in development projects, there was a model that relied on a limited amount of automation or human control in stages such as environment controls, software controls, versioning control, and inadequate approaches such as gitflow in applications and development approaches. This traditional approach, which combines waterfall and agile methodologies, was created with the obligation to comply with standards, strict requirements, project schedule and customer expectations, and could be insufficient to respond to changing needs.

As shown in Figure 1, the development in this method was tested by unit tests with code periodically uploaded to the development environment by team members for testing and debugging. The code was then transferred from local machines to remote servers using version management tools such as Subversion/git. Here, the advantages of using git and the advantages it provides were not seen due to the use of subversion instead of git or not applying the gitflow model. In the next stage, the code that is developed and made ready for publication by making bug fixes is labelled, versioned, and sent to the package repository with the continuous distribution/delivery tool. Here, although the upload to the repository is done with the continuous distribution/delivery tool, the processes for bugs and dependencies are not suitable for automation. In the next stage, uploading to environments is done manually, especially for field environments, test and production environments.

In this approach, configuration management was not very involved in the process, configuration management was not integrated very intensively until the product release phase, so it was not possible to keep the configuration of environments, software, change and all end-to-end tools integrated. In addition, since automation was not used intensively and project management tools such as Jira were not used effectively, tracking and controlling the information related to the publication, version, change, as well as frequent or even continuous delivery and keeping the environments up to date took a lot of time, so again a structure with low efficiency and feasibility was emerging.

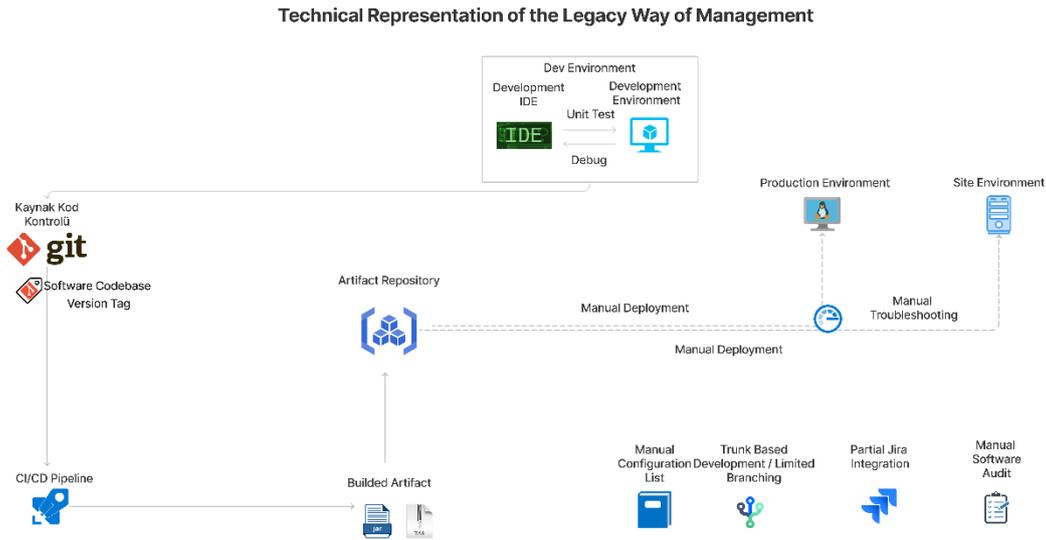

**Figure 1:** Traditional Method (DevOps and Agile Method Integrated Pre-SCM Structure)

Therefore, instead of the traditional software configuration management approach, an approach that integrates DevOps and Agile methods, as shown in Figure 3, has been developed and implemented in order to overcome productivity issues. This approach emphasises flexibility in development and management and Scrum, one of the agile methods, was applied. The Scrum approach is basically an agile method for managing and completing complex projects focussing on delivering valuable product increments in short time units (2 weeks) called sprints (Fowler & Fowler, 2019). The key features of Scrum include a collaborative and self-organising team, a flexible and incremental approach, regularly scheduled meetings, a task log list and sprint log list, and a focus on continuously delivering development, potentially a releaseable mini-product at the end of each sprint. The team also audits and adapts its process and product at the end of each sprint to improve its performance in the next sprint (Fowler & Fowler, 2019).

During the development process in our model, the configuration manager participated in task planning and daily scrum meetings, and version, change, documentation and development tasks were followed in accordance with this method. The project management tool Jira was used for communication and planning in this area. You can see the BPMN graph of this method in Figure 2.

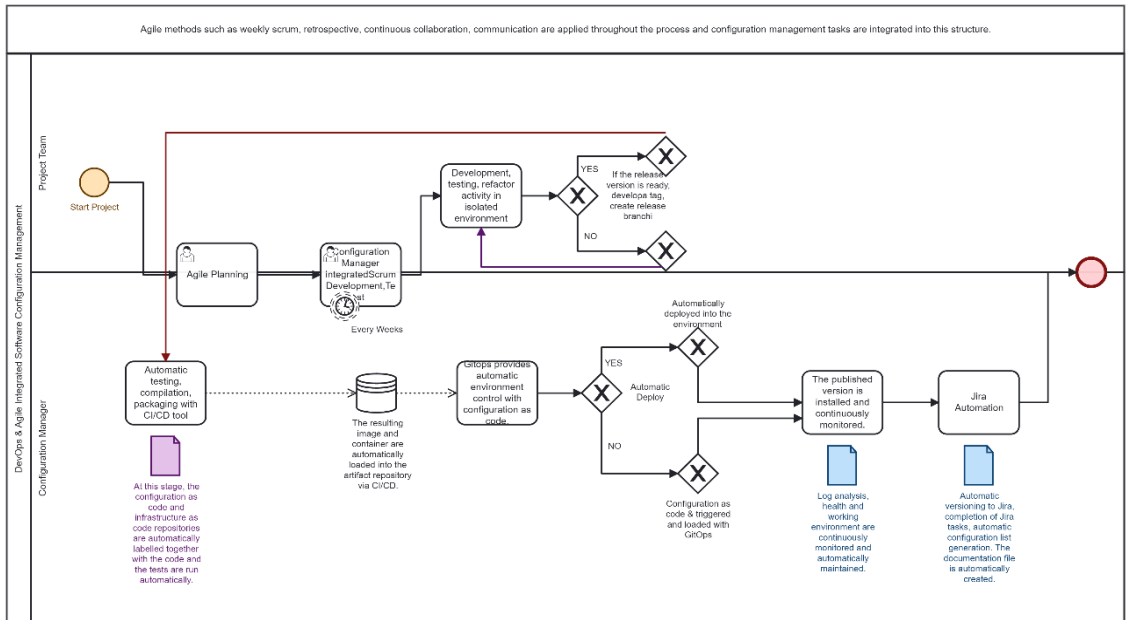

**Figure 2:** DevOps, Agile Method Integrated Software Configuration Management Approach BPMN Diagram

When we consider end-to-end, in the new model, a structure has been established that aims to continuously throw the code developed for the development team into the environment allocated for development, test and eliminate errors. After this stage, the code is sent from the local machine to a central git remote repository via git. Here, the gitflow approach is adopted and the principle that developers only send their code to the feature branches coming out of the develop branch is applied. The gitflow approach is essentially a branching model for Git that provides a robust framework for managing branches and releases in a software development project (Dwaraki, Seetharaman, Natarajan, & Wolf, 2015). It includes a master branch for production-ready code, a develop branch for current development, feature branches to implement new features, release branches to prepare for a release, and hotfix branches to quickly fix production issues. This model encourages developers to work on isolated feature branches and provides a clear process for managing releases and hotfixes, helping teams maintain a stable code base and avoid conflicts (Dwaraki, Seetharaman, Natarajan, & Wolf, 2015). With the flexibility provided by this structure, the code that is forwarded to the remote repository is moved to the develop branch when it is ready and labelled, allowing the configuration manager to plan the release. A model has been established to start the process with a single trigger via the configuration manager continuous integration/delivery (CI/CD) tool (Jenkins, TravisCI) for the code that is transmitted from the development environment to the remote repository and release.

In addition to the code developed in this structure, in the new approach, environment-based (production, test) configuration and infrastructure information is kept as code. This approach is called Configuration as Code and Infrastructure as Code (Lie, Sánchez-Gordón, & Colomo-Palacios, 2020). When this development code is triggered by the continuous integration and delivery tool, both development, configuration and infrastructure codes are automatically labelled on git with the relevant version number. It is compiled, prepared and automatically uploaded to the repository where it is kept version-based. In this way, the installable, workable software product consisting of the developed code is ready with the version number.

In this structure where the configuration repository manages the environments as code on Git, the version of the new software is updated in the relevant repository and automatically installed in the desired environment through the GitOps tool. Here, the condition of installation to the production environment is made dependent on the approval of the configuration manager in the configuration repository as code. In this way, while the software ready for publication is uploaded to the environment in an end-to-end, automated manner, its version is kept on git, a structure suitable for returning emerges, all changes are monitored and configuration control is ensured, especially in the change of the production environment. In addition, keeping both the configuration and the infrastructure as version-based and code has also provided a structure suitable for change, reversibility and control.

However, after this process, the continuous integration/infrastructure tool automatically goes and defines a version to the Jira tool thanks to the integration, maps each task developed on git to this version, and an automatic configuration list is created regarding which problems are solved in which version. This is important for standards in the defence industry.

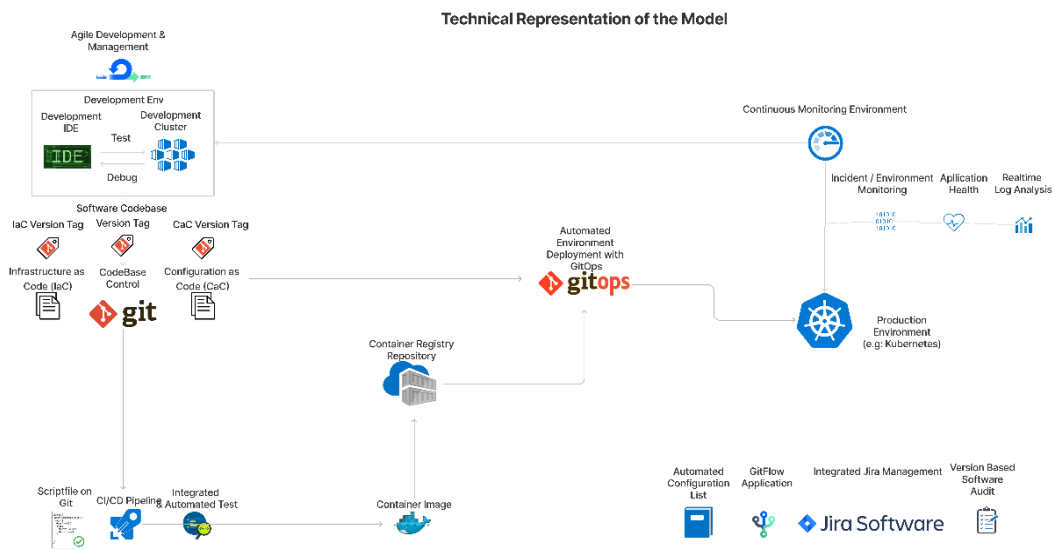

**Figure 3:** DevOps, Agile Method Integrated Software Configuration Management Approach

When we look at this model from the theoretical framework, it is seen that all activities of software configuration management (change management, release management, planning, definition, monitoring and control, teamwork, etc.) match with DevOps principles and activities (Hochbergs & Nilsson Sjödahl, 2020). Principles such as auditing, status control, change management are matched with the principles of monitoring, planning and continuous testing in the DevOps perspective (Lie, Sánchez-Gordón, & Colomo-Palacios, 2020). In addition, the integration of agile and DevOps methods in software configuration management provides many benefits and improvements.

The new model is strengthened by the advantages of both agile and DevOps methods. These advantages can be listed as faster deployment of releases and more frequent releases, increased control and monitoring, increased quality and reliability, improved collaboration and communication.

These advantages, such as faster deployment and more frequent releases, are made possible by the structure provided by DevOps and automation tools, continuous integration/delivery and GitOps tools and strategies. The increase in control and observation, and the increase in quality and reliability are made possible by the infrastructure-as-code and configuration-as-code method and keeping all these codes on git in a version-controlled manner. Thanks to the agile method, the benefit of continuous co-operation and communication is strengthened. The fact that the configuration manager works in integration with the team, follows his own tasks in integration with the team, handles, plans and manages the release planning in a managerial dimension, and does this openly thanks to the project management tool called Jira, and is involved in evaluations with daily meetings provides this benefit and strengthens the model.

## 4. Results

In this study, the configuration management approach in a defence industry company is examined before and after. While traditional approaches were applied in the previous model, an approach was developed by integrating methods such as automation, agility and DevOps in the new model. This approach was applied to improve and transform the existing process and model. The approach ensured that the basic principles of configuration management such as consistency, version management, release management, accuracy of records, baseline management, infrastructure and environment are recorded accurately and completely. This ensures the healthy operation of the software and environment, controlling the correct versions, ensuring communication between individuals and departments, defining the correct repository configuration and configuration elements and establishing appropriate structures. In addition, the control of configuration changes, code changes and infrastructure changes has resulted in a higher quality and secure infrastructure. With this method, the release time for the project decreased from 2 days to 3 hours, the release frequency increased up to 4 times, the code quality increased thanks to automated testing and integration, and an improvement of up to 1.5 times in the error rate was observed. In addition, there was a significant improvement in the time spent per broadcast and a more controlled structure was observed. Nevertheless, it was evaluated that there may be some limitations such as learning new technologies, human resources suitable for technology and application, and planning of the process.

**Acknowledgements.** The authors would like to thank HAVELSAN management, all departments and colleagues who contributed to the application, and Figen Şensoy, Configuration Management Team Leader, for her support.